\documentclass[conference]{IEEEtran}
\IEEEoverridecommandlockouts

\usepackage[T1]{fontenc}
\usepackage{amsmath,amsfonts}
\usepackage{array}
\usepackage[nolist,nohyperlinks]{acronym}
\usepackage{textcomp}
\usepackage{stfloats}
\usepackage{url}
\usepackage{verbatim}
\usepackage{graphicx}
\usepackage{cite}
\usepackage{xcolor}
\usepackage[english]{babel}
\usepackage{subfiles}
\usepackage{algorithm}
\usepackage{algpseudocode}
\usepackage{amsmath}
\usepackage{tabularx}
\usepackage{soul}

\usepackage{booktabs}
\usepackage{pdfpages}
\usepackage{hyperref}

\IEEEoverridecommandlockouts

\title{Towards Reliable Cloud VR over Wi-Fi 7: Leveraging MLO and TXOP Sharing}
\author{Hugo Sanz, Francesc Wilhelmi, Boris Bellalta   \\
\textit{Wireless Networking - Universitat Pompeu Fabra, Barcelona}\\
\texttt{\{hugo.sanz, francisco.wilhelmi, boris.bellalta\}@upf.edu}}
\date{}

\begin{document}

\IEEEpubid{\makebox[\columnwidth]{© 2026 IFIP\hfill}
\hspace{\columnsep}\makebox[\columnwidth]{}]}

\maketitle

\begin{abstract}
Cloud~\ac{VR} imposes stringent latency and reliability requirements due to the closed-loop interaction between \ac{DL} video streaming and \ac{UL} tracking updates. Although Wi-Fi~7 introduces \ac{MLO} to improve throughput and reduce latency, its adoption for latency-critical applications like Cloud~\ac{VR} is not straightforward. This paper investigates Cloud~\ac{VR} streaming over Wi-Fi~7 through a comprehensive cross-layer evaluation and proposes an application-aware \ac{TXS} scheduling policy that exploits the \ac{UL}/\ac{DL} coupling to jointly serve tracking and video traffic. To evaluate our mechanism, we introduce NeXus-VR, an open-source simulator that extends OMNeT++ to integrate the complete Cloud~\ac{VR} application pipeline and the Wi-Fi~7 MAC/PHY layers. Simulation results with multiple \ac{VR} users, \ac{OBSS} contention, and user mobility show that \ac{MLO} and the proposed \ac{TXS} scheduling policy provide complementary benefits. While \ac{MLO} improves transmission opportunities and robustness, the proposed Cloud \ac{VR}-aware \ac{TXS} scheduler preserves tracking freshness by reducing contention between \ac{UL} and \ac{DL} traffic. Together, they increase Cloud~\ac{VR} capacity by up to 2.25$\times$ over conventional single-link operation while consistently reducing latency and improving application-level quality of experience.
\end{abstract}


\acresetall

\section{Introduction}
\label{sec:intro}

\begin{figure}[t!]
    \centering    \includegraphics[width=0.9\columnwidth]{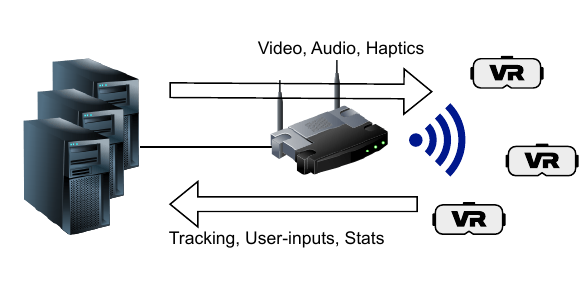}
    \caption{Cloud~VR Wi-Fi deployment.}
    \label{fig:VR-Wi-Fi_scenario}
\end{figure}

Immersive Cloud~\ac{VR} applications have experienced rapid growth in recent years, driven by advances in display technologies, graphics rendering, and edge computing~\cite{Akyildiz2022WirelessER,hou2017wireless}. In split-rendering Cloud~\ac{VR}, rendering is offloaded to an edge or cloud server, while lightweight headsets transmit tracking information and receive the rendered video stream over the wireless network (see Figure~\ref{fig:VR-Wi-Fi_scenario}). Since every video frame depends on the latest user pose, the \ac{UL} tracking and \ac{DL} video streams form a tightly coupled closed loop with stringent latency, jitter, and reliability requirements~\cite{wang2023simulator}. Delayed tracking or video delivery directly degrades the experience quality and even induces motion sickness. As wireless communications offer the necessary freedom of movement these kind of immersive applications, Wi-Fi (as a predominant wireless technology for indoor) is a grounded choice for running Cloud~\ac{VR}~\cite{carrascosa2025xr,BELLALTA2026}. However, the wireless access introduces a primary performance bottleneck~\cite{korneev2024model,michaelides2025lessons,BELLALTA2026}.

Recent IEEE 802.11 standards have progressively introduced mechanisms to make Wi-Fi support latency-sensitive applications. IEEE~802.11ax (Wi‑Fi~6) incorporates \ac{OFDMA} to improve efficiency in dense deployments, while IEEE~802.11be (Wi‑Fi~7) further increases transmission opportunities through wider channels, higher-order modulation, \ac{MLO}, multi-\ac{RU} allocation, and \ac{TXS}~\cite{khorov2018tutorial,perez2022time,deng2020ieee}. These enhancements improve throughput, latency, and scalability for \ac{XR} traffic~\cite{carrascosa2025xr,loginov2025scalability}. At the same time, the ongoing IEEE~802.11bn (Wi‑Fi~8) standardization is increasingly focused on enhancing reliability, latency predictability, and coordinated operation for emerging immersive and real-time applications~\cite{galati2024will,geraci2025wifitwentyfiveyearscounting}. Nevertheless, channel access remains contention-based, and interrelated \ac{UL} and \ac{DL} transmissions continue to expose Cloud~\ac{VR} applications to collisions, asymmetric delays, stale tracking information, and even \ac{MLO} starvation effects under certain operating conditions~\cite{carrascosa2025xr,michaelides2025lessons}.

Among recent advances in IEEE 802.11, \ac{TXS} unlocks flexibility at coordinating multiple transmissions within a common \ac{TXOP}, but intentionally leaves the scheduling policy unspecified. 
We propose an application-aware TXS policy specifically designed for split rendering Cloud VR. Although not specifically targeting Cloud~\ac{VR}, the work in~\cite{gokhale2023vitals} is closely related, as it explores a similar \ac{TXS} approach to support the stringent latency requirements of Tactile Internet applications.

This paper investigates the performance of Cloud~\ac{VR} over Wi-Fi~7 through a comprehensive cross-layer evaluation. We propose a Cloud \ac{VR}-aware \ac{TXS} scheduler that jointly coordinates \ac{UL} tracking and \ac{DL} video transmissions within each shared \ac{TXOP}, reducing channel contention while preserving tracking freshness. To evaluate its benefits, we have developed NeXus-VR,\footnote{The source code of NeXus-VR is openly available at \url{https://github.com/hugosanzbenito/NeXus-VR}.}, an open-source simulator that extends OMNeT++ to simulate a complete Cloud~\ac{VR} application pipeline over Wi-Fi~7 networks. Unlike conventional network-centric studies, the proposed framework enables the simultaneous evaluation of Wi-Fi performance metrics and application-level Cloud~\ac{VR} metrics, allowing us to quantify how wireless impairments translate into end-user performance. The main contributions of this paper are summarized as follows:
\begin{itemize}
\item \textbf{Cloud~\ac{VR}-aware \ac{TXS} scheduling.}
We propose a scheduling policy that jointly serves \ac{UL} tracking and \ac{DL} video traffic within each shared \ac{TXOP}. By exploiting \ac{OFDMA} for concurrent \ac{UL} transmissions, the proposed \ac{TXS} scheduler reduces contention, improves channel utilization, and preserves tracking freshness.
\item \textbf{Cross-layer Cloud~\ac{VR} simulation framework.}
We introduce NeXus-VR, a new OMNeT++ framework that jointly models the complete split-rendering Cloud~\ac{VR} pipeline and Wi-Fi~7 MAC/PHY operation, enabling the simultaneous evaluation of networking and application-level performance metrics.
\item \textbf{Joint evaluation of \ac{MLO} and \ac{TXS}.}
We quantify the individual and combined impact of Wi-Fi~7 \ac{MLO} and Cloud~\ac{VR}-aware \ac{TXS} scheduling on Cloud~\ac{VR} performance under realistic operating conditions.
\item \textbf{Comprehensive performance assessment.}
We evaluate single-\ac{AP} multi-user deployments, \ac{OBSS} contention, and user mobility, showing that \ac{MLO} and Cloud~\ac{VR}-aware \ac{TXS} scheduling provide complementary benefits that substantially improve capacity, latency, robustness, and overall Cloud~\ac{VR} application-level metrics.
\end{itemize}

The remainder of this paper is organized as follows. Section~\ref{sec:enhancements} presents the Cloud~\ac{VR} architecture, the Wi-Fi~7 communication model, and the proposed application-aware \ac{TXS} scheduling policy. Section~\ref{sec:sysmodel} describes the cross-layer simulation framework and evaluation methodology. Section~\ref{sec:performance} presents the performance evaluation. Finally, Section~\ref{sec:conclusions} concludes the paper.


\section{Cloud VR Streaming over Wi-Fi}
\label{sec:enhancements}

\begin{figure*}[t!]
    \centering
    \includegraphics[width=\linewidth]{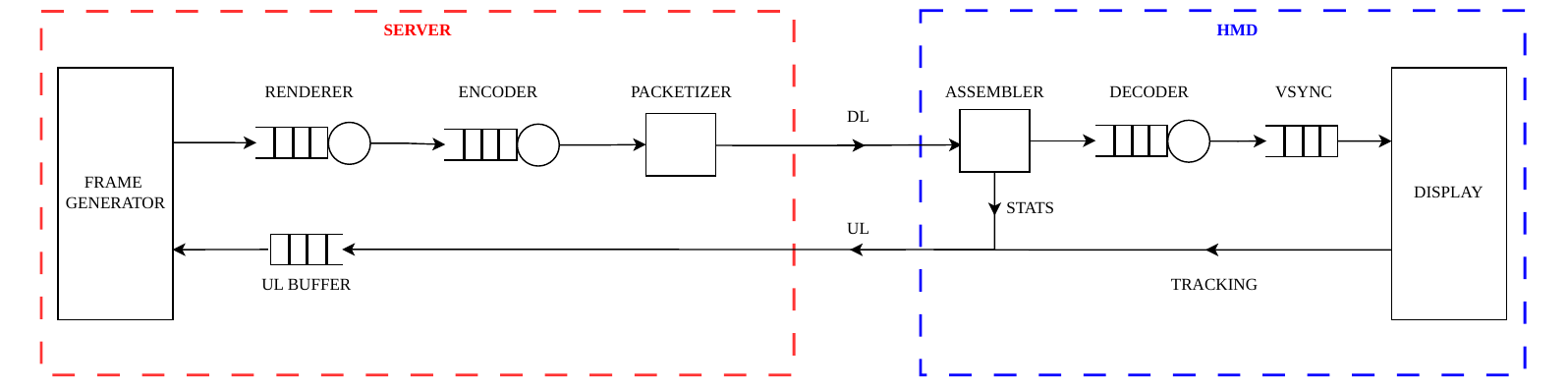}
    \caption{End-to-end Cloud~VR processing pipeline.}
    \label{fig:ServerVR}
\end{figure*}

This section presents the Cloud~\ac{VR} architecture considered throughout the paper. We first describe the Cloud~\ac{VR} pipeline and its interaction with the wireless network. After that, we introduce Wi-Fi~7 \ac{MLO} and discuss its benefits for low-latency traffic. Finally, we present the proposed Cloud~\ac{VR}-aware \ac{TXS} scheduler, which exploits the flexibility offered by Wi-Fi~7 to better align wireless resource allocation with the requirements of immersive \ac{VR} applications.

\subsection{Cloud VR Pipeline}

We consider a split-rendering Cloud~\ac{VR} architecture in which scene rendering is performed on a remote server while user interaction and visualization take place on a wireless \ac{VR} headset connected through a Wi-Fi~7 \ac{AP}. The server and the headset exchange two tightly coupled traffic streams: a high-rate \ac{DL} video stream that delivers the rendered video frames and a low-rate \ac{UL} stream that transmits tracking information and application feedback. Unlike conventional video streaming, Cloud~\ac{VR} traffic exhibits a strong dependency between both traffic directions. Every rendered frame is generated from the latest tracking information available at the rendering server. Consequently, the freshness of \ac{UL} tracking packets directly impacts the quality of the \ac{DL} video stream.

Figure~\ref{fig:ServerVR} illustrates the considered end-to-end Cloud~\ac{VR} pipeline. Tracking information generated by the headset is periodically transmitted to the rendering server, where it drives the rendering process. Rendered frames are encoded, packetized, and delivered through the Wi-Fi network. At the headset, packets are reassembled into frames, decoded, and synchronized with the display before being presented to the user. Frames that cannot be reconstructed before their playback deadline are discarded. This processing chain introduces multiple latency components, including rendering, encoding, wireless transmission, reassembly, decoding, and display synchronization. While application processing delays depend on the rendering platform, wireless latency is largely determined by the operation of the Wi-Fi network. Efficient coordination between the application and communication layers is therefore essential to minimize the \ac{MTP} latency.

\subsection{Multi-Link Operation}

Wi-Fi~7 introduces \ac{MLO}, allowing \acp{MLD} to communicate over multiple independent frequency bands. In this work, we consider the \ac{STR} \ac{MLMR} mode, in which independent radio interfaces can simultaneously transmit and receive over different links~\cite{carrascosa2023understanding}. By exploiting multiple links in parallel, \ac{STR} \ac{MLMR} increases transmission opportunities, reduces channel access delays, and improves robustness against channel fluctuations and external interference.

For Cloud~\ac{VR}, \ac{STR} \ac{MLMR} is particularly attractive because both the \ac{DL} video stream and the \ac{UL} tracking traffic are highly latency-sensitive. Multiple independent links reduce queuing delays and increase the probability that latency-critical packets can be transmitted promptly, especially under network congestion or when one of the links temporarily experiences poor channel conditions. However, complexity arises because, without a proper scheduler, \ac{MLO} treats traffic as if it was generated independently, thus \ac{UL} tracking packets and \ac{DL} video frames compete for transmission opportunities. Consequently, although \ac{STR} \ac{MLMR} increases the available transmission resources, it does not exploit the intrinsic dependency between both traffic directions in Cloud~\ac{VR}. The proposed Cloud~\ac{VR}-aware \ac{TXS} scheduler addresses this limitation by jointly coordinating \ac{UL} and \ac{DL} transmissions within each shared \ac{TXOP}.

\subsection{Cloud VR-aware TXOP Sharing Scheduling}

\begin{figure*}[t!]
    \centering
    \includegraphics[width=\linewidth]{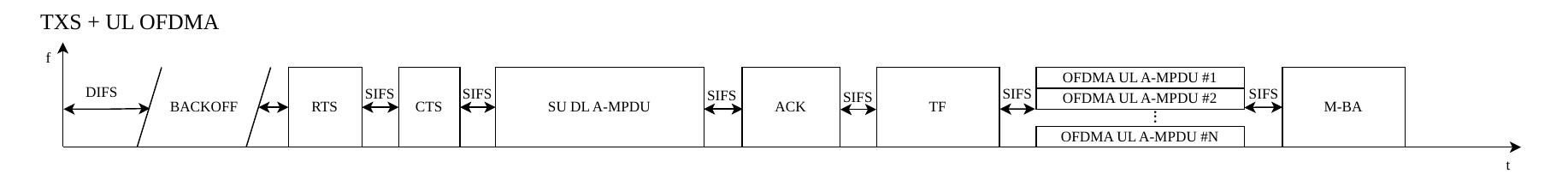}
    \caption{Proposed Cloud-VR-aware TXS transmission. In the figure, TF stands for Trigger Frame, and M-BA for Multi-STA Block ACK.}
    \label{fig:placeholder}
\end{figure*}

Although IEEE~802.11be introduces the \ac{TXS} mechanism to improve channel utilization, the standard intentionally leaves the scheduling policy unspecified. This flexibility enables application-aware schedulers that coordinate transmissions according to higher-layer requirements. In Cloud~\ac{VR}, \ac{UL} tracking packets and \ac{DL} video frames are conventionally scheduled independently, requiring separate channel contentions despite their tight temporal relationship. As a result, latency-critical tracking updates may experience additional channel access delays, unnecessarily increasing the \ac{MTP} latency.

To address this limitation, we propose a Cloud~\ac{VR}-aware \ac{TXS} scheduler that jointly coordinates \ac{DL} video traffic and \ac{UL} tracking packets within each shared \ac{TXOP}. After acquiring a \ac{TXOP} through the standard \ac{EDCA} channel access procedure, the \ac{AP} first transmits the buffered \ac{DL} video traffic using \ac{A-MPDU} aggregation. Rather than releasing the medium after completing the \ac{DL} transmission, the \ac{AP} exploits the remaining \ac{TXOP} duration to schedule a Trigger-based \ac{UL} \ac{OFDMA} transmission, allowing the selected \acp{STA} to simultaneously deliver their tracking updates, as illustrated in Figure~\ref{fig:placeholder}. Consequently, \ac{UL} tracking packets no longer contend independently for channel access, eliminating additional channel contention before their transmission. Finally, the \ac{AP} acknowledges the received packets using a \ac{M-BA} frame.

For \ac{UL} transmissions, the scheduler allocates a minimum of one 20~MHz \ac{RU} to each \ac{STA}. This design choice provides sufficient capacity for the low-rate tracking traffic while allowing an 80~MHz channel to simultaneously accommodate up to four \acp{STA}. When the number of active \acp{STA} exceeds the available \acp{RU}, the scheduler distributes transmission opportunities across successive \acp{TXOP} using a round-robin policy. Buffer-aware scheduling is unnecessary because tracking packets are generated periodically with nearly constant size, while collecting real-time buffer occupancy information from all \acp{STA} would introduce additional signaling overhead with negligible scheduling gains. \ac{DL} \ac{OFDMA} is intentionally not employed because the \ac{AP} buffer typically contains highly asymmetric amounts of queued video data for different \acp{STA}, leading to inefficient \ac{RU} allocation and underutilization of the available transmission resources.

During frame aggregation, the scheduler continuously estimates the transmission duration, accounting for payload, MAC/PHY overheads, acknowledgements, control exchanges, and inter-frame spaces. Aggregation stops as soon as adding another \ac{DL} packet, or the Trigger-based \ac{UL} transmission, would exceed the maximum IEEE~802.11be \ac{PPDU} duration, leaving the remaining packets buffered for the next \ac{TXOP}.

Overall, the proposed scheduler exploits the flexibility provided by IEEE~802.11be \ac{TXS} to jointly schedule \ac{DL} video and \ac{UL} tracking traffic within a single transmission opportunity. By eliminating an independent channel contention for every tracking update, it reduces wireless access delay, lowers the \ac{MTP} latency, and improves the scalability of Cloud~\ac{VR} networks without requiring any modification to the IEEE~802.11be \ac{MAC} protocol.

\section{Evaluation Methodology}
\label{sec:sysmodel}

To evaluate the proposed scheduler, we developed a custom OMNeT++-based cross-layer simulator. The simulator implements a complete split-rendering Cloud~\ac{VR} system over Wi-Fi~7, including the application pipeline and the corresponding PHY/MAC functionalities, enabling the joint evaluation of networking and application-level performance. Next, we describe the methodology adopted to develop the simulator, including the Cloud~\ac{VR} application model, the wireless network model, and the performance metrics.

\subsection{Cloud VR Application Model}

We model a split-rendering Cloud~\ac{VR} application following the operation of ALVR,\footnote{\url{https://github.com/alvr-org/ALVR}} where a remote rendering server generates the video stream while periodically receiving head-tracking updates from the wireless headset. The application pipeline comprises rendering, video encoding, packetization, wireless transmission, packet reassembly, video decoding, and display synchronization. \ac{DL} traffic consists of a constant-bitrate video stream, while \ac{UL} traffic carries head-tracking updates and application feedback. Additional details on the considered traffic characteristics are provided in~\cite{BELLALTA2026}.

\begin{table*}[t!!!]
\centering
\caption{Cloud~\ac{VR} and Network Performance Metrics}
\label{tab:KPIs}
\begin{tabularx}{\textwidth}{lX}
\toprule
\textbf{Metric} & \textbf{Description} \\
\midrule

Video Frame Round-Trip Time (VF-RTT) &
\ac{VF-RTT}, defined as the elapsed time between the generation of a video frame at the rendering server and the reception of the corresponding application statistics packet. This metric represents the latency introduced by the wireless network. \\
\midrule
Motion-to-Photon (MTP) &
\ac{MTP} latency is defined as the end-to-end delay between a user's physical movement and the display of the corresponding video frame. In this work, it is measured as the pose-to-render-to-photon latency, i.e., from the sampling of a pose at the headset to the display of the frame rendered based on that pose, including uplink and downlink transmission as well as server- and client-side processing delays.\\
\midrule
Frames without Tracking (NoTrack) &
Percentage of rendered video frames generated without fresh tracking information owing to delayed \ac{UL} updates. \\
\midrule
Consecutive stale-tracking frames (Av. Burst) &
Average number of consecutive stale-tracking frames. This metric quantifies the temporal persistence of tracking freezes and is strongly correlated with user discomfort. \\
\midrule
Timeout Discards (TOut D.) &
Percentage of video frames discarded because not all packets belonging to the frame were received before the reassembly timeout expired. \\
\midrule
Decoder Discards (Dec. D.) &
Percentage of video frames discarded owing to decoder queue overflow. \\
\midrule
Vsync Discards (Vsync D.) &
Percentage of decoded video frames discarded because the display synchronization buffer overflowed. \\
\midrule
Frame Loss Ratio (FLR) &
\ac{FLR}, defined as the percentage of generated video frames discarded at any stage of the application pipeline, including TOut D., Dec. D., and Vsync D.\\
\midrule
Not Ready to Project (NRP) &
Percentage of display refresh events in which no decoded frame was available for presentation, potentially resulting in a visible freeze or stutter. \\
\bottomrule
\end{tabularx}
\end{table*}

\subsection{Wireless Network Model}

We consider an indoor Cloud~\ac{VR} deployment comprising $N$ rendering servers connected to a Wi-Fi~7 \ac{AP} through a 10~Gbps Ethernet backbone, which serves $N$ wireless \ac{VR} headsets. Each server is paired with one headset, generating \ac{DL} video traffic and receiving \ac{UL} tracking information. The \ac{AP} is located at the center of a $10\times10$~m indoor area. The wireless network operates either on a single 5~GHz link in \ac{SLO} mode or on two 80~MHz links in the 5 and 6~GHz bands in \ac{MLO} mode. In all cases, channel access is based on the EDCA Best Effort access category.

Signal propagation follows the indoor TMB model~\cite{adame2019tmb}, adapted to both frequency bands. The path loss between the \ac{AP} and a headset at a distance $d$ is computed as
\begin{equation}
\mathrm{PL}(d)=
L_0+
10\gamma\log_{10}(d)+
k\overline{W}d,
\nonumber
\end{equation}
where $L_0$~dB represents the reference path loss at 1~m, $\gamma$ is the path loss exponent, $k$~dB is the wall attenuation factor per meter, and $W$~walls/m denotes the wall density.

Link adaptation is modeled through received-power thresholds that dynamically select the highest feasible \ac{MCS} assuming perfect channel estimation. Frames received below the carrier-sense threshold are discarded. PHY transmission rates follow the IEEE~802.11be specification, while an independent \ac{PER} is applied to every transmitted \ac{MPDU}, allowing individual subframes within an \ac{A-MPDU} to be corrupted independently.

To evaluate the impact of neighboring Wi-Fi networks, we incorporate a simple \ac{OBSS} contention model. Whenever a link is ready to start the backoff procedure, the simulator checks whether the channel is occupied by an external transmission. This occurs with probability $p_{\mathrm{OBSS}}$. If the channel is busy, the node defers channel access for a duration $T_{\mathrm{OBSS}}$, where $T_{\mathrm{OBSS}}$ is exponentially distributed, after which the channel state is evaluated again. Otherwise, the standard backoff procedure starts immediately. For \ac{MLO}, the OBSS contention process is modeled independently on each link, such that the occupancy of one channel does not affect the probability of the other being occupied.

User mobility follows the \ac{RWP} model implemented in the OMNeT++ INET framework. Each \ac{STA} moves within the deployment area at a uniformly distributed speed between 0.5 and 1.0~m/s. Upon reaching a destination, the \ac{STA} pauses for a uniformly distributed interval between 1 and 2~s before selecting a new waypoint. Mobility is restricted to the horizontal plane, while the vertical coordinate remains fixed throughout the simulation.


\subsection{Performance Metrics}

The considered Cloud~\ac{VR} system is evaluated using both networking and application-level metrics. While conventional networking metrics quantify the efficiency of the wireless channel, application-level metrics capture the impact of delays on the Cloud~\ac{VR} user experience. Together, they provide a comprehensive characterization of the overall system performance. Table~\ref{tab:KPIs} summarizes the metrics considered throughout the evaluation.

Among the application-level metrics, particular attention is paid to the \ac{MTP} latency, which is one of the primary determinants of the perceived \ac{QoE} in immersive applications and is explicitly recognized in ITU-T recommendations for Cloud~\ac{VR} services~\cite{ITU_F74614}. Since the wireless network constitutes a major component of the end-to-end latency budget, we also evaluate its contribution through the \ac{VF-RTT}. Based on our previous experimental study~\cite{casasnovas2025nestvr}, maintaining an average \ac{VF-RTT} below approximately 33~ms was found to consistently provide a good interactive Cloud~\ac{VR} experience. The same study showed that the \ac{FLR} should remain below 1\% to avoid noticeable degradation in user experience. Accordingly, we use these values as representative reference thresholds throughout the evaluation, while application-level metrics are reported only for illustration and are not used as evaluation criteria.

\section{Performance Evaluation}
\label{sec:performance}

\begin{table}[t!!!]
\centering
\caption{Evaluation Parameters}
\label{tab:parameters}
\begin{tabular}{ll}
\toprule
\textbf{Parameter} & \textbf{Value} \\
\midrule

\multicolumn{2}{l}{\textit{Cloud~\ac{VR} Application}}\\
\midrule
Video bitrate & 100 Mbps \\
Frame rate & 90 FPS \\
Traffic model & ALVR~\cite{BELLALTA2026} \\
Rendering service rate ($\exp$) & 200 FPS \\
Encoder service rate ($\exp$) & 200 FPS \\
Decoder service rate ($\exp$) & 200 FPS \\
Display refresh rate & 90 FPS \\
Encoder and Decoder buffer size & 4 frames \\
Reassembly timeout & 50 ms \\
\midrule
\multicolumn{2}{l}{\textit{Network}}\\
\midrule
Ethernet Link (between VR servers and AP) & 10 Gbps \\ 
Deployment area & $10\times10$ m \\
Frequency bands & 5 and 6 GHz \\
Channel width & 80 MHz \\
Propagation model & TMB \\
$L_0$ & 54.12 dB \\
$\gamma$ & 2.06067 \\
$k$ & 5.25 dB \\
$\overline{W}$ & 0.1467 walls/m \\
Transmit power & 20 dBm \\
CCA threshold & $-82$ dBm \\
Supported MCSs & 0--13 \\
Link adaptation & Received-power thresholds \\
Packet error rate (PER) & 10\% \\
$p_{\mathrm{OBSS}}$ & 0.1 \\
Mean $T_{\mathrm{OBSS}}$  ($exp$) & 2 ms \\
\midrule
\multicolumn{2}{l}{\textit{MAC and Mobility}}\\
\midrule
Maximum TXOP duration & 5.484 ms \\
Minimum RU bandwidth & 20 MHz \\
Mobility model & Random Waypoint \\
Walking speed & Uniform [0.5, 1.0] m/s \\
Pause time & Uniform [1, 2] s \\
Mobility plane & 2-D ($z$ fixed) \\
\bottomrule
\end{tabular}
\end{table}

This section evaluates the proposed Cloud \ac{VR} system from both networking and application perspectives. We first assess its scalability by comparing conventional transmission and \ac{TXS} under \ac{SLO} and \ac{MLO} operation, and then evaluate their robustness to external contention and user mobility. Throughout the evaluation, conducted using the parameters in Table~\ref{tab:parameters}, each simulation point is evaluated over 20 independent runs of 50~s, with Wi-Fi and application-level \ac{VR} metrics jointly considered to assess overall system performance.

\subsection{Performance of Cloud VR-aware \ac{TXS} scheduling under SLO}
\label{subsection:RQ1}

We first evaluate the proposed \ac{TXS} scheduling in a \ac{SLO} Wi-Fi deployment. The objective is to quantify the benefits of jointly scheduling \ac{UL} tracking and \ac{DL} video within each \ac{TXOP} on the maximum number of \ac{VR} users that can be supported while meeting the stringent latency requirements of immersive applications. All users are statically located 5~m from the \ac{AP}.

\begin{figure}[h!]
    \centering
    \makebox[\linewidth][c]{\includegraphics[width=1.05\linewidth]{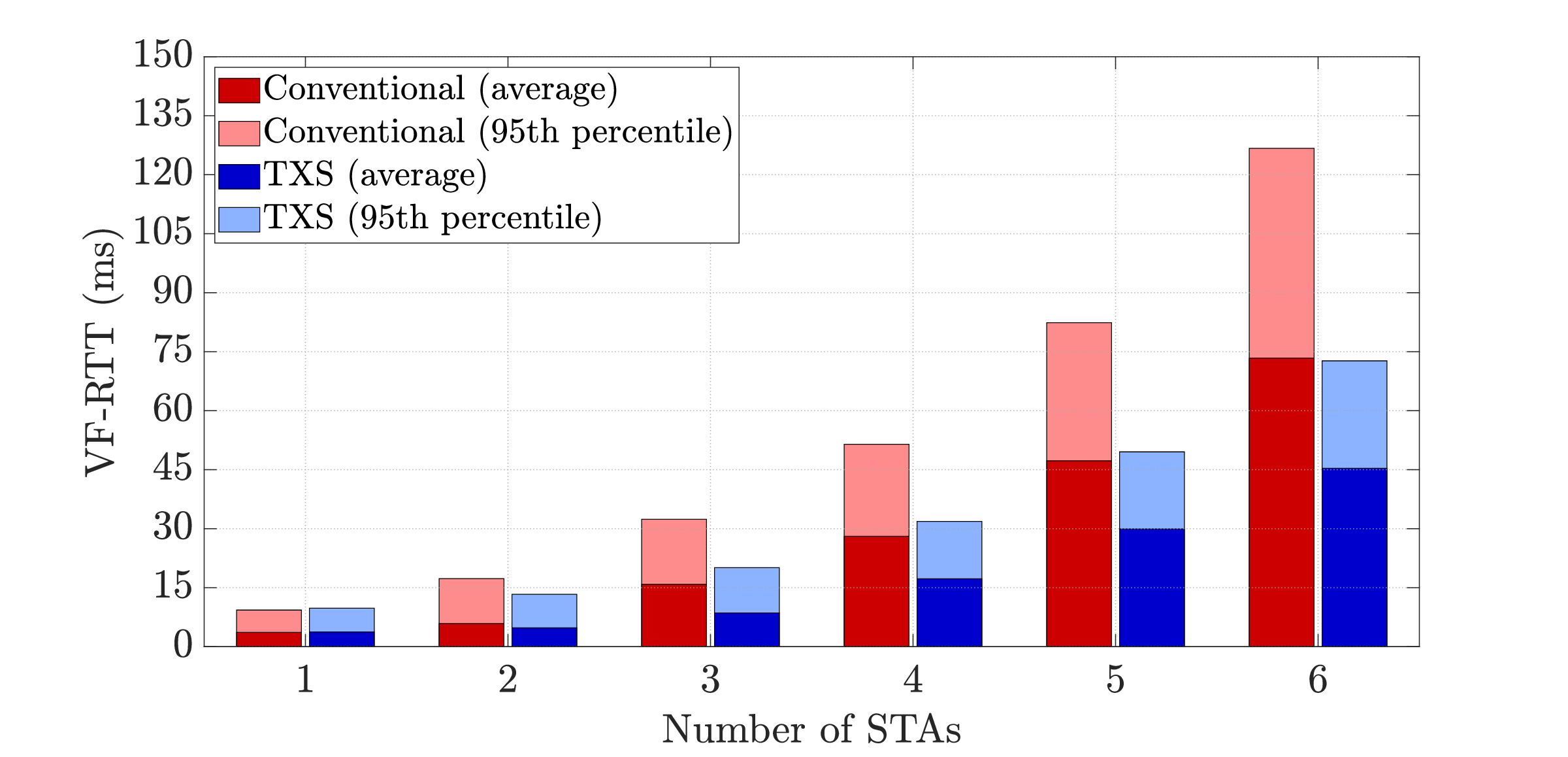}}
    \caption{VF-RTT under SLO.}
    \label{fig:VFRTT_1}
\end{figure}

Figure~\ref{fig:VFRTT_1} compares the \ac{VF-RTT} under \ac{SLO} for conventional transmission (without \ac{TXS}) and the proposed \ac{TXS} scheduling. Both approaches achieve average \ac{VF-RTT} values close to 5~ms with one or two users, although \ac{TXS} reduces latency by 18\% with two users. As the number of users increases, conventional transmission becomes rapidly contention-limited, with average \ac{VF-RTT} reaching approximately 30~ms with four users and exceeding 70~ms with six users. In contrast, \ac{TXS} jointly schedules \ac{UL} tracking and \ac{DL} video within each \ac{TXOP}, keeping average \ac{VF-RTT} around 30~ms with five users and 45~ms with six users, with substantially lower variability. However, the \ac{FLR} remains above 1\% at five users, limiting the supported capacity to four users.

Table~\ref{tab:comparison_RQ1} provides further insight into the application-level performance. Under the conventional \ac{SLO} transmission (without \ac{TXS}), the percentage of video frames generated using outdated tracking information increases from less than 4\% with two users to over 60\% with five users, severely degrading the Cloud~\ac{VR} experience. In contrast, enabling \ac{TXS} under \ac{SLO} keeps this percentage below 6\% even in the six-user scenario by prioritizing the delivery of tracking updates before video transmission. Moreover, for both conventional transmission and \ac{TXS}, the percentage of frames missing the timeout deadline (\emph{TOut D.}) starts to increase noticeably from five users onward, indicating that the network approaches saturation.

\begin{table*}[t!!]
\centering
\caption{Performance metrics comparison between conventional \ac{SLO} operation (without \ac{TXS}) and \ac{SLO} with \ac{TXS}}
\vspace{0.15cm}
{
\setlength{\tabcolsep}{4pt}
\noindent\makebox[\textwidth][c]{
\resizebox{1\linewidth}{!}{
\begin{tabular}{|c||cc|cc|cc||cc|cc|cc|cc||cc|}
\hline
 & \multicolumn{2}{c|}{Avg. MTP (ms)} & \multicolumn{2}{c|}{NoTrack. (\%)} & \multicolumn{2}{c||}{Avg. Burst} & \multicolumn{2}{c|}{Dec. D. (\%)} & \multicolumn{2}{c|}{Vsync D. (\%)} & \multicolumn{2}{c|}{TOut D. (\%)} & \multicolumn{2}{c||}{FLR (\%)} & \multicolumn{2}{c|}{NRP (\%)} \\ \hline
STAs & Conv. & TXS & Conv. & TXS & Conv. & TXS & Conv. & TXS & Conv. & TXS & Conv. & TXS & Conv. & TXS & Conv. & TXS \\
\hline
1 & 46.48 & 47.20 & 1.14  & 0.09 & 0.01 & 0.00 & 0.00 & 0.00 & 0.00 & 0.00 & 0.00  & 0.00  & 0.00  & 0.00  & 0.87  & 0.87  \\
2 & 56.87 & 51.91 & 3.39  & 0.47 & 0.04 & 0.00 & 0.00 & 0.00 & 0.01 & 0.02 & 0.00  & 0.00  & 0.01  & 0.02  & 0.70  & 0.72  \\
3 & 66.74 & 56.91 & 17.85 & 0.52 & 0.22 & 0.01 & 0.01 & 0.00 & 0.05 & 0.04 & 0.02  & 0.00  & 0.08  & 0.04  & 0.40  & 0.36  \\
4 & 76.57 & 64.61 & 43.09 & 0.54 & 0.76 & 0.01 & 0.06 & 0.01 & 0.22 & 0.12 & 0.02  & 0.01  & 0.30  & 0.14  & 0.52  & 0.34  \\
5 & 93.41 & 72.41 & 62.81 & 1.21 & 1.69 & 0.01 & 2.27 & 1.50 & 0.62 & 0.36 & 4.86  & 3.70  & 7.75  & 5.56  & 14.41 & 12.22  \\
6 & 115.58 & 82.70 & 75.34 & 5.88 & 3.05 & 0.06 & 6.67 & 4.65 & 0.23 & 0.31 & 16.87 & 9.89  & 23.77 & 14.84 & 24.39 & 15.48 \\ \hline
\end{tabular}
}
}
}
\label{tab:comparison_RQ1}
\end{table*}

\vspace{0.25cm}
\noindent\textbf{Takeaway:}
Conventional \ac{SLO} operation is fundamentally limited by the independent channel contention between \ac{UL} tracking and \ac{DL} video traffic, restricting the system to four concurrent \ac{VR} users. Enabling \ac{TXS} under \ac{SLO} largely eliminates stale tracking by jointly scheduling both traffic directions within each \ac{TXOP} and consistently improves the evaluated application-level metrics. However, at five users, the \ac{FLR} remains above the 1\% threshold despite the improved \ac{VF-RTT}, limiting the supported capacity to four users. For one to four users, \ac{TXS} provides consistent improvements across all evaluated application-level metrics.

\subsection{Scalability Gains with Multi-Link Operation}
\label{subsection:RQ2}

We next evaluate the proposed \ac{TXS} mechanism under \ac{MLO}, assessing how the additional transmission opportunities provided by \ac{MLO} interact with the joint scheduling of \ac{UL} tracking and \ac{DL} video. The objective is to quantify the maximum \ac{VR} capacity achievable when combining \ac{MLO} with \ac{TXS}. All users are statically located 5~m from the \ac{AP}.

\begin{figure}[h!]
    \centering
    \makebox[\linewidth][c]{\includegraphics[width=1.05\linewidth]{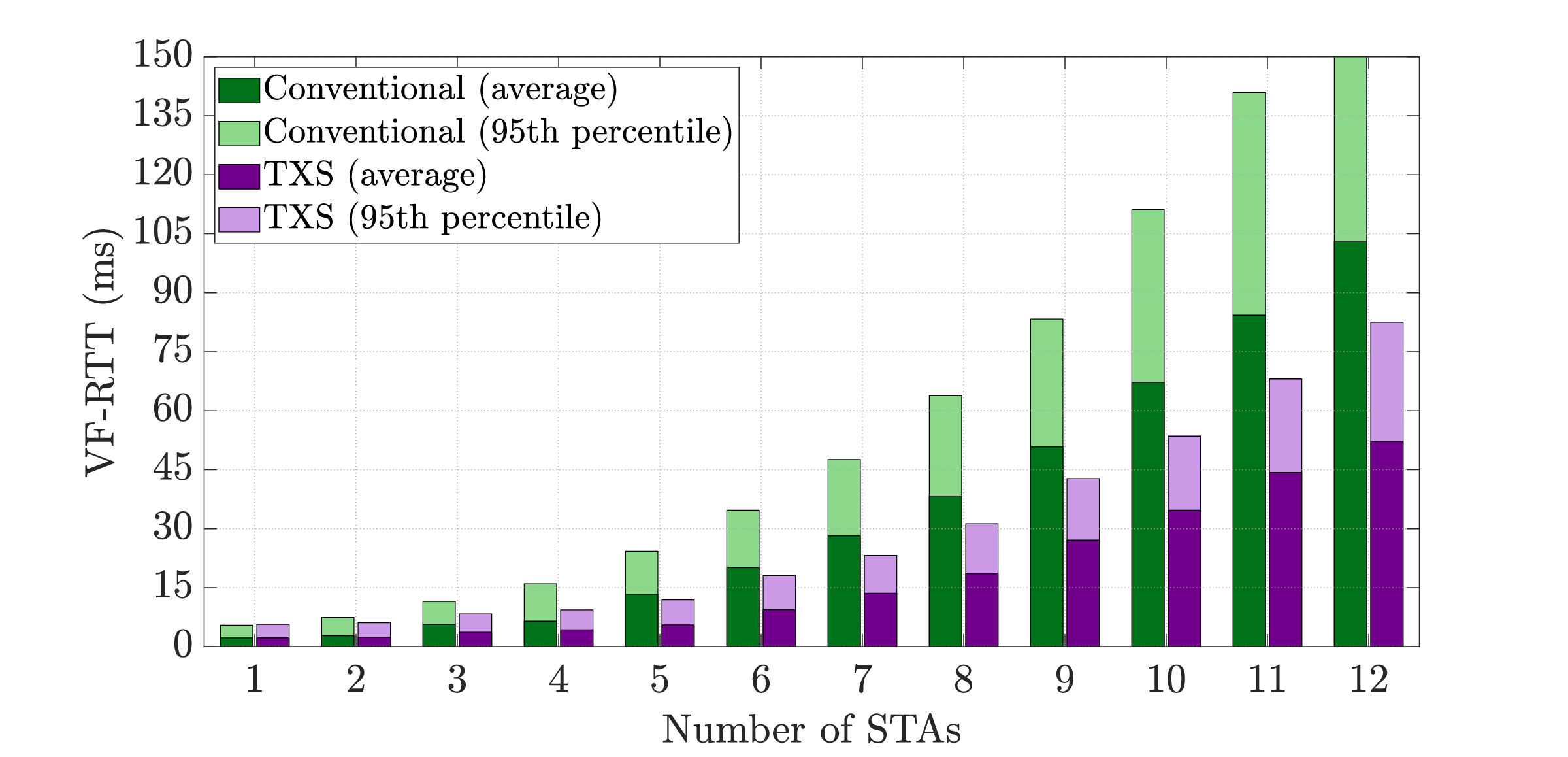}}
    \caption{VF-RTT under MLO.}
    \label{fig:VFRTT_2}
\end{figure}

Fig.~\ref{fig:VFRTT_2} presents the evolution of the \ac{VF-RTT} as the number of associated users increases. As expected, \ac{MLO} substantially increases network capacity by providing multiple independent transmission opportunities. Nevertheless, enabling \ac{TXS} consistently improves performance across the entire evaluated range. For two to four users, \ac{TXS} reduces the average \ac{VF-RTT} by 15--35\%, despite the network operating well below saturation. As contention increases, the performance gap widens, and conventional \ac{MLO} without \ac{TXS} exceeds the 33~ms average \ac{VF-RTT} threshold beyond seven users. In contrast, \ac{MLO} with \ac{TXS} maintains latency below this threshold for up to nine users. Beyond this point, both approaches become constrained by channel capacity, although \ac{TXS} continues to provide significantly lower latency and lower latency variability.

The application-level metrics reported in Table~\ref{tab:comparison_RQ2} confirm these observations. Under conventional \ac{MLO} operation, stale tracking again represents the primary scalability bottleneck, exceeding 10\% with only five users and rapidly increasing thereafter. In contrast, enabling \ac{TXS} under \ac{MLO} almost completely eliminates this limitation, maintaining stale-tracking rates below 10\% up to eleven users. As in the \ac{SLO} scenario, the dominant limitation eventually shifts from tracking freshness to frame losses and playback deadlines once the wireless network approaches saturation.

\begin{table*}[t!!]
\centering
\caption{Performance metrics comparison between conventional \ac{MLO} operation (without \ac{TXS}) and \ac{MLO} with \ac{TXS}}
\vspace{0.15cm}
{
\setlength{\tabcolsep}{4pt}
\noindent\makebox[\textwidth][c]{
\resizebox{1\linewidth}{!}{
\begin{tabular}{|c||cc|cc|cc||cc|cc|cc|cc||cc|}
\hline
 & \multicolumn{2}{c|}{Avg. MTP (ms)} & \multicolumn{2}{c|}{NoTrack. (\%)} & \multicolumn{2}{c||}{Avg. Burst} & \multicolumn{2}{c|}{Dec. D. (\%)} & \multicolumn{2}{c|}{Vsync D. (\%)} & \multicolumn{2}{c|}{TOut D. (\%)} & \multicolumn{2}{c||}{FLR (\%)} & \multicolumn{2}{c|}{NRP (\%)} \\ \hline
STAs & Conv. & TXS & Conv. & TXS & Conv. & TXS & Conv. & TXS & Conv. & TXS & Conv. & TXS & Conv. & TXS & Conv. & TXS \\
\hline
1  & 41.26 & 38.97 & 0.07  & 0.09  & 0.00 & 0.00 & 0.00 & 0.00 & 0.00 & 0.00 & 0.00  & 0.00  & 0.00  & 0.00  & 0.87  & 0.85  \\
2  & 45.22 & 40.28 & 0.13  & 0.04  & 0.00 & 0.00 & 0.00 & 0.00 & 0.00 & 0.00 & 0.00  & 0.00  & 0.00  & 0.00  & 0.66  & 0.67  \\
3  & 45.88 & 40.33 & 0.44  & 0.11  & 0.00 & 0.00 & 0.00 & 0.00 & 0.00 & 0.00 & 0.00  & 0.00  & 0.00  & 0.00  & 0.28  & 0.30  \\
4  & 46.02 & 42.04 & 1.41  & 0.03  & 0.01 & 0.00 & 0.00 & 0.00 & 0.00 & 0.00 & 0.00  & 0.00  & 0.00  & 0.00  & 0.17  & 0.51  \\
5  & 57.57 & 45.31 & 10.37 & 0.04  & 0.12 & 0.00 & 0.00 & 0.00 & 0.00 & 0.00 & 0.00  & 0.00  & 0.00  & 0.00  & 0.62  & 0.21  \\
6  & 66.73 & 51.40 & 25.91 & 0.02  & 0.35 & 0.00 & 0.00 & 0.00 & 0.01 & 0.00 & 0.00  & 0.00  & 0.01  & 0.00  & 0.61  & 0.60  \\
7  & 73.87 & 57.07 & 42.82 & 0.07  & 0.75 & 0.00 & 0.01 & 0.00 & 0.03 & 0.01 & 0.00  & 0.00  & 0.04  & 0.01  & 0.74  & 0.66  \\
8  & 83.59 & 63.18 & 55.29 & 0.27  & 1.24 & 0.00 & 0.18 & 0.01 & 0.16 & 0.04 & 0.08  & 0.00  & 0.42  & 0.05  & 0.97  & 0.54  \\
9  & 94.57 & 69.25 & 64.77 & 1.05  & 1.84 & 0.01 & 1.24 & 0.13 & 0.34 & 0.18 & 1.03  & 0.04  & 2.61  & 0.36  & 3.23  & 1.00  \\
10 & 107.64 & 74.90 & 72.43 & 2.36  & 2.63 & 0.02 & 5.47 & 0.94 & 0.25 & 0.26 & 9.54  & 0.72  & 15.26 & 1.92  & 15.92 & 2.56  \\
11 & 125.01 & 81.35 & 78.10 & 6.12  & 3.56 & 0.07 & 8.75 & 4.56 & 0.12 & 0.15 & 23.78 & 6.93  & 32.65 & 11.64 & 33.36 & 12.28 \\
12 & 146.00 & 89.42 & 82.62 & 11.69 & 4.75 & 0.13 & 9.31 & 8.12 & 0.13 & 0.09 & 42.39 & 17.72 & 51.83 & 25.93 & 52.29 & 26.53 \\ \hline
\end{tabular}
}
}
}
\label{tab:comparison_RQ2}
\end{table*}

\vspace{0.25cm}
\noindent\textbf{Takeaway:}
\ac{MLO} increases Cloud~\ac{VR} capacity by providing additional transmission opportunities, while \ac{TXS} further reduces the contention between \ac{UL} tracking and \ac{DL} video by jointly scheduling both traffic directions within each \ac{TXOP}. Combined, they increase the supported capacity from seven users with conventional \ac{MLO} operation to nine with \ac{TXS}, corresponding to a 1.28$\times$ improvement over the Wi-Fi~7 baseline, and a 2.25$\times$ increase with respect to conventional \ac{SLO} operation.

\subsection{Impact of \ac{OBSS} Contention}
\label{subsection:RQ3}

We next evaluate the robustness of \ac{TXS} under increasing external interference. To isolate the effect of \ac{OBSS}, the number of \ac{VR} users is fixed to 2 while the probability of external channel occupation is progressively increased from 0 to 40\%. Four configurations are compared: conventional transmission and \ac{TXS} under both \ac{SLO} and \ac{MLO}. All users are statically located 5~m from the \ac{AP}.

\begin{figure}[t!!!]
    \centering
    \makebox[\linewidth][c]{\includegraphics[width=1.05\linewidth]{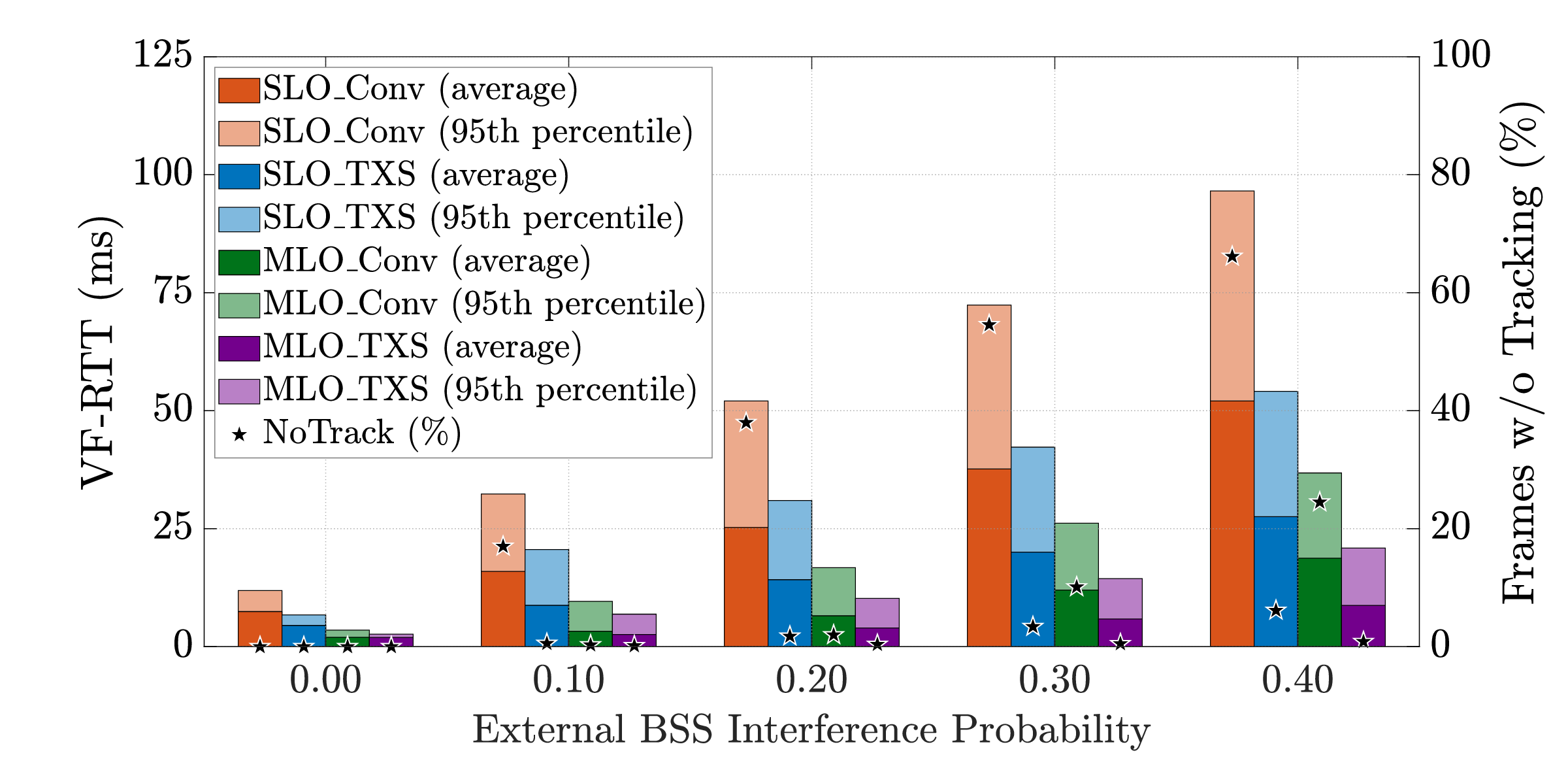}}
    \caption{\ac{VF-RTT} and stale-tracking frames under increasing \ac{OBSS} interference.}
    \label{fig:vfrtt_obss}
\end{figure}

Fig.~\ref{fig:vfrtt_obss} reports the average \ac{VF-RTT} as the \ac{OBSS} activity increases. External contention has a pronounced impact on \ac{SLO} deployments, where every deferred transmission blocks the only available wireless link. Consequently, conventional \ac{SLO} operation experiences a continuous latency increase, with the average \ac{VF-RTT} growing from approximately 5~ms without interference to nearly 50~ms when 40\% of transmission opportunities are blocked. Although enabling \ac{TXS} under \ac{SLO} mitigates this degradation by reducing internal contention, its performance remains fundamentally constrained by the availability of a single wireless channel.

\ac{MLO} fundamentally changes this behavior by allowing packets to be transmitted over an alternative link whenever one becomes temporarily unavailable. As a result, both \ac{MLO} configurations exhibit substantially higher resilience to external contention. Nevertheless, enabling \ac{TXS} under \ac{MLO} consistently achieves the lowest latency by combining link diversity with the joint scheduling of \ac{UL} tracking and \ac{DL} video, maintaining an average \ac{VF-RTT} below 10~ms and an \ac{FLR} below 1\%, even under the highest evaluated \ac{OBSS} contention level.

\vspace{0.25cm}
\noindent\textbf{Takeaway:}
\ac{TXS} primarily addresses internal contention by jointly scheduling \ac{UL} tracking and \ac{DL} video traffic, whereas \ac{MLO} mitigates external contention through link diversity. For the two-user scenario considered, their combination provides the highest robustness, maintaining \ac{VF-RTT} below 10~ms even when up to 40\% of transmission opportunities are blocked by neighboring Wi-Fi networks.

\subsection{Impact of User Mobility}
\label{subsection:RQ4}

Finally, we evaluate the proposed Cloud \ac{VR} system under user mobility. Unlike the previous static scenarios, users move according to the \ac{RWP} model within a $10\times10$~m room, with the \ac{AP} located at its center. This configuration captures the channel fluctuations experienced in realistic \ac{VR} deployments, where users continuously change their position and, consequently, their wireless channel conditions.

\begin{figure}[t]
    \centering
    \makebox[\linewidth][c]{\includegraphics[width=1.05\linewidth]{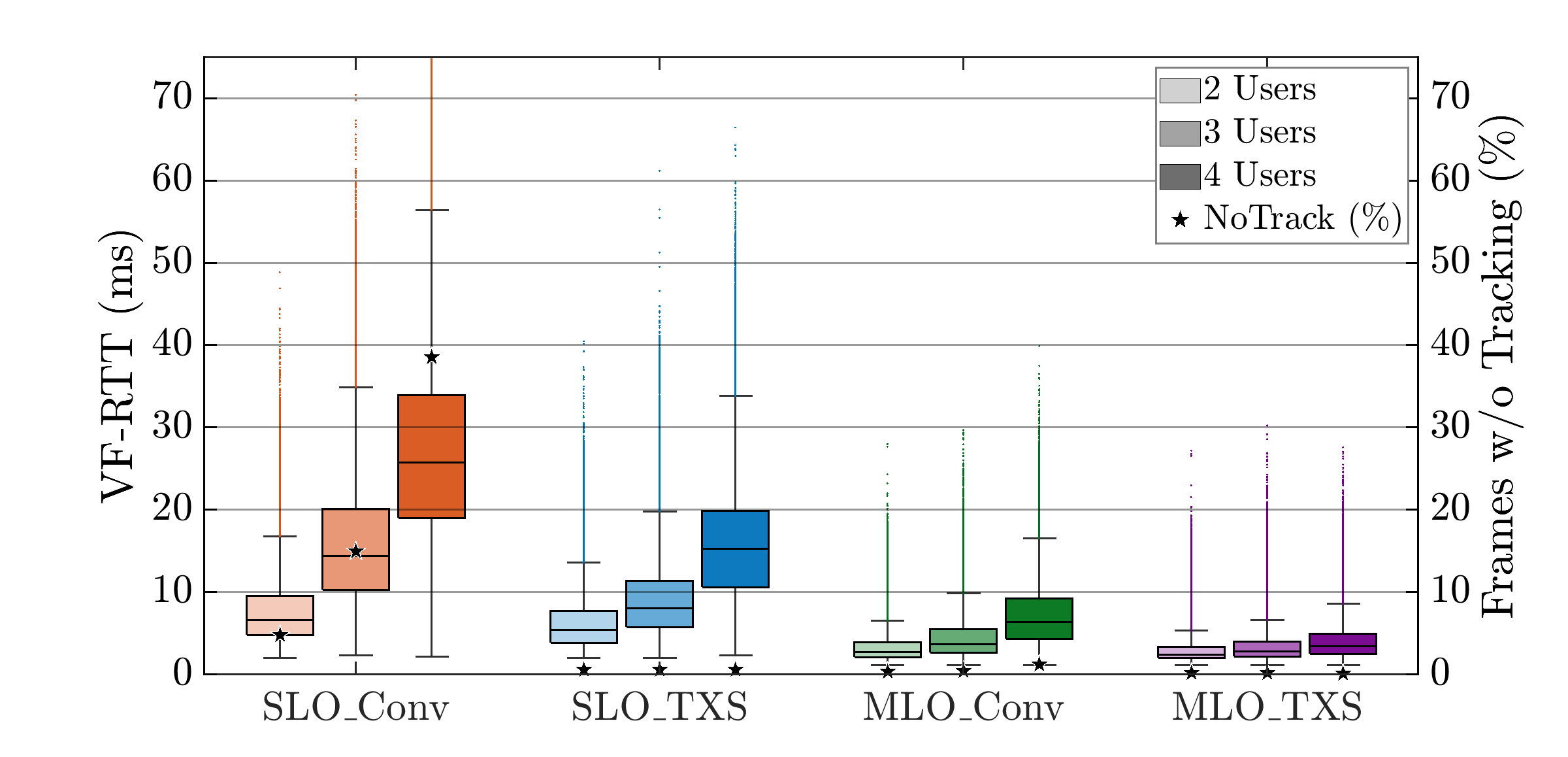}}
    \caption{\ac{VF-RTT} and stale-tracking frames under user mobility.}
    \label{fig:vfrtt_mobility}
\end{figure}

Figure~\ref{fig:vfrtt_mobility} presents the distribution of the \ac{VF-RTT} together with the percentage of video frames generated using outdated tracking information. Although users spend most of their time close to the \ac{AP}, occasional movements towards the room boundaries reduce the selected \ac{MCS} and increase transmission airtime. Owing to the well-known Wi-Fi performance anomaly, these transient channel degradations not only increase the latency of the affected user but also penalize all other users by occupying the shared medium for longer periods. Consequently, conventional transmission experiences a rapid increase in both latency and stale-tracking frames, particularly under \ac{SLO}, where approximately 15\% of frames are generated using outdated tracking information with three users and almost 40\% with four users.

Enabling \ac{TXS} effectively mitigates these channel fluctuations by prioritizing \ac{UL} tracking transmissions, maintaining stale-tracking rates close to zero throughout the evaluated scenarios while consistently reducing the \ac{VF-RTT}. \ac{MLO} further improves robustness by exploiting link diversity, reducing the impact of temporary channel degradation. Consequently, the combination of \ac{MLO} and \ac{TXS} achieves the lowest latency, the smallest variability, and the highest tracking freshness across all evaluated scenarios. 

\vspace{0.25cm}
\noindent\textbf{Takeaway:}
Mobility primarily degrades \ac{VR} performance by increasing transmission airtime during periods of poor channel quality. Due to the Wi-Fi performance anomaly, users experiencing poor channel conditions also penalize the performance of all other users by occupying the medium for longer periods. \ac{TXS} mitigates this effect by prioritizing tracking freshness through the joint scheduling of \ac{UL} and \ac{DL} traffic, while \ac{MLO} absorbs channel occupancy fluctuations through link diversity. Together, they provide the most robust solution, maintaining consistently low latency and fresh tracking under realistic user mobility.

\section{Conclusions}
\label{sec:conclusions}

This paper studied a cross-layer Cloud~\ac{VR} system for split-rendering over Wi-Fi~7, exploiting \ac{TXS} to jointly schedule latency-critical \ac{UL} tracking and high-throughput \ac{DL} video. A simulation framework integrating the Wi-Fi protocol stack and Cloud~\ac{VR} application pipeline was developed to evaluate the system from both networking and application perspectives. Results showed that conventional Wi-Fi~7 operation is limited by independent \ac{UL}/\ac{DL} contention, making stale tracking a key scalability bottleneck. \ac{TXS} mitigates this issue by jointly scheduling both traffic directions within each \ac{TXOP}, reducing latency and improving tracking freshness. While it consistently improves application-level metrics, the \ac{FLR} remains above 1\% at five users under \ac{SLO}, keeping the supported capacity at four users. Combined with \ac{MLO}, however, \ac{TXS} supports up to 9 simultaneous users, corresponding to 1.28$\times$ the capacity of conventional \ac{MLO} and 2.25$\times$ that of conventional \ac{SLO}. The system also remains robust under external contention and user mobility. Overall, the results highlight the benefits of joint application--MAC design, with application-aware \ac{TXS} complementing Wi-Fi~7 mechanisms such as \ac{MLO} to improve Cloud~\ac{VR} performance and scalability.

\section*{Acknowledgments}

This work was supported by the following projects: TRUE Wi-Fi PID2024-155470NB-I00/MICIU/AEI/10.13039/501100011033/FEDER,UE; REALM (GA 101298050, funded by the EU); ICREA Academia 2024 program (2024 ICREA 00077/AGAUR/Department of Research and Universities of the Government of Catalonia); and MdM CEX2021-001195-M/MICIU/AEI/10.13039/501100011033. The views and opinions expressed are those of the authors and do not necessarily reflect those of the European Union; neither the European Union nor the granting authority can be held responsible for them.  

\vspace{0.5cm}

\noindent (C) IFIP, 2026. This is the author's version of the work. It is posted here by permission of IFIP for your personal use. Not for redistribution. The definitive version was published in IFIP WMNC 2026.


\bibliographystyle{IEEEtran}
\bibliography{References}

\begin{acronym}[AAAAAAAAA]  
 \acro{ED}{energy detection}
 \acro{PD}{packet detection}
 \acro{CSMA/CA}{carrier sense multiple access/collision avoidance}
 \acro{CCA}{clear channel assessment}
 \acro{CW}{contention window}
 \acro{CF-End}{control frame-end}
 \acro{2D}{two-dimensional}
 \acro{3D}{three-dimensional}
 \acro{3G}{third generation}
 \acro{3GPP}{Third Generation Partnership Project}
 \acro{4G}{fourth generation}
 \acro{5G}{fifth generation}
 \acro{6G}{sixth generation}
 \acro{5GC}{5G core network}
 \acro{AAA}{authentication, authorisation and accounting}
 \acro{A-BFT}{association beamforming training}
 \acro{ABM}{agent-based model}
 \acro{ABR}{adaptive bit rate}
 \acro{ABS}{almost blank subframe}
 \acro{AC}{access category}
 \acro{ACIR}{adjacent channel interference rejection ratio}
 \acro{ACK}{acknowledgment}
 \acro{ACL}{allowed CSG list}
 \acro{ACLR}{adjacent channel leakage ratio}
 \acro{ACPR}{adjacent channel power ratio}
 \acro{ACS}{adjacent channel selectivity}
 \acro{ADC}{analog-to-digital converter}
 \acro{ADSL}{asymmetric digital subscriber line}
 \acro{AEE}{area energy efficiency}
 \acro{AF}{amplify-and-forward}
 \acro{AGCH}{access grant channel}
 \acro{AGG}{aggressor cell}
 \acro{AH}{authentication header}
 \acro{AI}{artificial intelligence}
 \acro{AKA}{authentication and key agreement}
 \acro{AMC}{adaptive modulation and coding}
 \acro{ANN}{artificial neural network}
 \acro{ANR}{automatic neighbour relation}
 \acro{AoA}{angle of arrival}
 \acro{AoD}{angle of departure}
 \acro{AP}{access point}
 \acro{APC}{area power consumption}
 \acro{API}{application programming interface}
 \acro{APP}{a posteriori probability}
 \acro{AR}{augmented reality}
 \acro{ARIMA}{autoregressive integrated moving average}
 \acro{ARQ}{automatic repeat request}
 \acro{AS}{access stratum}
 \acro{ASE}{area spectral efficiency}
 \acro{ASM}{advanced sleep mode}
 \acro{ASN}{access service network}
 \acro{ATM}{asynchronous transfer mode}
 \acro{ATSC}{Advanced Television Systems Committee}
 \acro{AUC}{authentication centre}
 \acro{AWGN}{additive white gaussian noise}
 \acro{BA}{block acknowledgement}
 \acro{BACK}{block acknowledgement}
 \acro{BAR}{block acknowledgement request}
 \acro{BB}{baseband}
 \acro{BBU}{baseband unit}
 \acro{BCCH}{broadcast control channel}
 \acro{BCH}{broadcast channel}
 \acro{BCJR}{Bahl-Cocke-Jelinek-Raviv} 
 \acro{BE}{best effort}
 \acro{BEB}{binary exponential backoff}
 \acro{BER}{bit error rate}
 \acro{BFT}{beamforming training}
 \acro{BLER}{block error rate}
 \acro{BPSK}{binary phase-shift keying}
 \acro{BR}{bit rate}
 \acro{BRP}{beam refinement protocol}
 \acro{BS}{base station}
 \acro{BSC}{base station controller}
 \acro{BSIC}{base station identity code}
 \acro{BSP}{binary space partitioning}
 \acro{BSRP}{buffer status report poll}
 \acro{BSS}{blind source separation}
 \acro{BTS}{base transceiver station}
 \acro{BWP}{bandwidth part}
 \acro{CA}{carrier aggregation}
 \acro{CAC}{call admission control}
 \acro{CaCo}{carrier component}
 \acro{CAPEX}{capital expenditure}
 \acro{capex}{capital expenses}
 \acro{CAS}{cluster angular spread}
 \acro{CATV}{community antenna television}
 \acro{CAZAC}{constant amplitude zero auto-correlation}
 \acro{CBR}{constant bitrate}
 \acro{CC}{component carrier}
 \acro{CCCH}{common control channel}
 \acro{CCDF}{complementary cumulative distribution function}
 \acro{CCE}{control channel element}
 \acro{CCO}{coverage and capacity optimisation}
 \acro{CCPCH}{common control physical channel}
 \acro{CCRS}{coordinated and cooperative relay system}
 \acro{CCTrCH}{coded composite transport channel}
 \acro{CDF}{cumulative distribution function}
 \acro{CDMA}{code division multiple access}
\acro{CDR}{call detail record}
 \acro{CDS}{cluster delay spread}
 \acro{CESM}{capacity effective SINR mapping}
 \acro{CO$_{2e}$}{carbon dioxide equivalent}
 \acro{CFI}{control format indicator}
 \acro{CFL}{Courant-Friedrichs-Lewy}
 \acro{CGI}{cell global identity}
 \acro{CID}{connection identifier}
 \acro{CIF}{carrier indicator field}
 \acro{CIO}{cell individual offset}
 \acro{CIR}{channel impulse response}
 \acro{CNN}{convolutional neural network}
 \acro{CMF}{cumulative mass function}
 \acro{C-MIMO}{cooperative MIMO}
 \acro{CN}{core network}
 \acro{COC}{cell outage compensation}
 \acro{COD}{cell outage detection}
 \acro{CoMP}{coordinated multi-point}
 \acro{ConvLSTM}{convolutional LSTM}
 \acro{CP}{cycle prefix}
 \acro{CPC}{cognitive pilot channel}
 \acro{CPCH}{common packet channel}
 \acro{CPE}{customer premises equipment}
 \acro{CPICH}{common pilot channel}
 \acro{CPRI}{common public radio interface}
 \acro{CPU}{central processing unit}
 \acro{CQI}{channel quality indicator}
 \acro{CR}{cognitive radio}
 \acro{CRAN}{centralized radio access network} 
 \acro{C-RAN}{cloud radio access network} 
 \acro{CRC}{cyclic redundancy check}
 \acro{CRE}{cell range expansion}
 \acro{C-RNTI}{cell radio network temporary identifier}
 \acro{CRP}{cell re-selection parameter}
 \acro{CRS}{cell-specific reference symbol}
 \acro{CRT}{cell re-selection threshold}
 \acro{CSCC}{common spectrum coordination channel}
 \acro{CSG ID}{closed subscriber group ID}
 \acro{CSG}{closed subscriber group}
 \acro{CSI}{channel state information}
 \acro{CSIR}{receiver-side channel state information}
 \acro{CSI-RS}{channel state information-reference signals}
 \acro{CSO}{cell selection offset}
 \acro{Co-TDMA}{coordinated time division multiple access}
 \acro{Co-RTWT}{coordinated restricted target wake up time}
 \acro{CTCH}{common traffic channel}
 \acro{CTS}{clear-to-send} 
 \acro{CU}{central unit}
 \acro{CV}{cross-validation}
 \acro{CWiND}{Centre for Wireless Network Design}
 \acro{D2D}{device to device}
 \acro{DAB}{digital audio broadcasting}
 \acro{DAC}{digital-to-analog converter}
 \acro{DAS}{distributed antenna system}
 \acro{dB}{decibel}
 \acro{dBi}{isotropic-decibel}
 \acro{DBPS}{data bits per symbol}
 \acro{DC}{direct current}
 \acro{DCCH}{dedicated control channel}
 \acro{DCF}{distributed coordination function}
 \acro{DCH}{dedicated channel}
 \acro{DC-HSPA}{dual-carrier high speed packet access}
 \acro{DCI}{downlink control information}
 \acro{DCM}{directional channel model}
 \acro{DCP}{dirty-paper coding}
 \acro{DCS}{digital communication system}
 \acro{DD}{decoder discards}
 \acro{DECT}{digital enhanced cordless telecommunication}
 \acro{DeNB}{donor eNodeB}
 \acro{DFP}{dynamic frequency planning}
 \acro{DFS}{dynamic frequency selection}
 \acro{DFT}{discrete Fourier transform}
 \acro{DFTS}{discrete Fourier transform spread}
 \acro{DHCP}{dynamic host control protocol}
 \acro{DIFS}{DCF interframe space}
 \acro{DL}{downlink}
 \acro{DMC}{dense multi-path components}
 \acro{DMF}{demodulate-and-forward}
 \acro{DMT}{diversity and multiplexing tradeoff}
  \acro{DNN}{deep neural network} 
 \acro{DoA}{direction-of-arrival}
 \acro{DoD}{direction-of-departure}
 \acro{DoS}{denial of service}
 \acro{DPCCH}{dedicated physical control channel}
 \acro{DPDCH}{dedicated physical data channel}
 \acro{D-QDCR}{distributed QoS-based dynamic channel reservation}
 \acro{DQL}{deep Q-learning}
\acro{DRAN}{distributed radio access network}
 \acro{DRS}{discovery reference signal}
 \acro{DRL}{deep reinforcement learning}
 \acro{DRX}{discontinuous reception}
 \acro{DS}{down stream}
 \acro{DSA}{dynamic spectrum access}
 \acro{DSCH}{downlink shared channel}
 \acro{DSL}{digital subscriber line}
 \acro{DSLAM}{digital subscriber line access multiplexer}
 \acro{DSO}{dynamic subband operation}
 \acro{DSP}{digital signal processor}
 \acro{DT}{drive test}
 \acro{DTCH}{dedicated traffic channel}
 \acro{DTX}{discontinuous transmission}
   \acro{DU}{distributed unit}
 \acro{DVB}{digital video broadcasting}
 \acro{DXF}{drawing interchange format}
 \acro{E2E}{end-to-end}
 \acro{EAGCH}{enhanced uplink absolute grant channel}
 \acro{EA}{evolutionary algorithm}
 \acro{EAP}{extensible authentication protocol}
 \acro{EC}{evolutionary computing}
 \acro{ECGI}{evolved cell global identifier}
 \acro{ECR}{energy consumption ratio}
 \acro{ECRM}{effective code rate map}
 \acro{EDCA}{enhanced distributed channel access}
 \acro{MU-EDCA}{multi-user \ac{EDCA}}
 \acro{EDCH}{enhanced dedicated channel}
 \acro{EE}{energy efficiency}
 \acro{EESM}{exponential effective SINR mapping}
 \acro{EF}{estimate-and-forward}
 \acro{EGC}{equal gain combining}
 \acro{EHICH}{EDCH HARQ indicator channel}
 \acro{EHT}{extremely high throughput}
 \acro{eICIC}{enhanced intercell interference coordination}
 \acro{EIR}{equipment identity register}
 \acro{EIRP}{effective isotropic radiated power}
 \acro{ELF}{evolutionary learning of fuzzy rules}
 \acro{eMBB}{enhanced mobile broadband}
 \acro{eMLSR}{enhanced multi-link single radio}
  \acro{EMR}{Electromagnetic-Radiation}
 \acro{EMS}{enhanced messaging service}
 \acro{eNB}{evolved NodeB}
 \acro{eNodeB}{evolved NodeB}
 \acro{EoA}{elevation of arrival}
 \acro{EoD}{elevation of departure}
 \acro{EPB}{equal path-loss boundary}
 \acro{EPC}{evolved packet core}
 \acro{EPDCCH}{enhanced physical downlink control channel}
 \acro{EPLMN}{equivalent PLMN}
 \acro{EPS}{evolved packet system}
 \acro{ERAB}{eUTRAN radio access bearer}
 \acro{ERGC}{enhanced uplink relative grant channel}
 \acro{ERTPS}{extended real time polling service}
 \acro{ESB}{equal downlink receive signal strength boundary}
 \acro{ESF}{even subframe}
 \acro{ESP}{encapsulating security payload}
 \acro{ESS}{extended service set}
 \acro{ETSI}{European telecommunications standards institute}
 \acro{E-UTRA}{evolved UTRA}
 \acro{EU}{European Union}
 \acro{EUTRAN}{evolved UTRAN}
 \acro{EVDO}{evolution-data optimised}
 \acro{FACCH}{fast associated control channel}
 \acro{FACH}{forward access channel}
 \acro{FAP}{femtocell access point}
 \acro{FARL}{fuzzy assisted reinforcement learning}
 \acro{FCC}{Federal Communications Commission}
 \acro{FCCH}{frequency-correlation channel}
 \acro{FCFS}{first-come first-served}
 \acro{FCH}{frame control header}
 \acro{FCI}{failure cell ID}
 \acro{FD}{frequency-domain}
 \acro{FDD}{frequency division duplexing}
 \acro{FDM}{frequency division multiplexing}
 \acro{FDMA}{frequency division multiple access}
 \acro{FDTD}{finite-difference time-domain}
 \acro{FE}{front-end}
 \acro{FeMBMS}{further evolved multimedia broadcast multicast service}
 \acro{FER}{frame error rate}
 \acro{FFR}{fractional frequency reuse}
 \acro{FFRS}{fractional frequency reuse scheme}
 \acro{FFT}{fast Fourier transform}
 \acro{FFU}{flexible frequency usage}
 \acro{FGW}{femtocell gateway}
 \acro{FIFO}{first-in first-out}
 \acro{FIS}{fuzzy inference system}
 \acro{FLR}{frame loss ratio}
 \acro{FMC}{fixed mobile convergence}
 \acro{FPC}{fractional power control}
 \acro{FPS}{frames per second}
 \acro{FPGA}{field-programmable gate array}
 \acro{FRS}{frequency reuse scheme}
 \acro{FSPL}{free space path loss}
 \acro{FTP}{file transfer protocol}
 \acro{FTTx}{fiber to the x}
 \acro{FUSC}{full usage of subchannels}
 \acro{GA}{genetic algorithm}
 \acro{GAN} {generative adversarial network}
 \acro{GANC}{generic access network controller}
 \acro{GBR}{guaranteed bitrate}
 \acro{GCI}{global cell identity}
 \acro{GERAN}{GSM edge radio access network}
 \acro{GGSN}{gateway GPRS support node}
 \acro{GHG}{greenhouse gas}
 \acro{GMSC}{gateway mobile switching centre}
 \acro{gNB}{next generation NodeB}
 \acro{GNN}{graph neural network}
 \acro{GCN}{graph convolutional network}
 \acro{GNSS}{global navigation satellite system}
 \acro{GP}{genetic programming}
 \acro{GPON}{Gigabit passive optical network}
 \acro{GPP}{general purpose processor}
 \acro{GPRS}{general packet radio service}
 \acro{GPS}{global positioning system}
 \acro{GPU}{graphics processing unit}
 \acro{GRU}{gated recurrent unit}
 \acro{GSCM}{geometry-based stochastic channel models}
 \acro{GSMA}{global system for mobile communications association}
 \acro{GSM}{global system for mobile communication}
 \acro{GTD}{geometry theory of diffraction}
 \acro{GTP}{GPRS tunnel protocol}
 \acro{GTP-U}{GPRS tunnel protocol - user plane}
 \acro{GW}{gateway}
 \acro{HA}{hybrid access}
 \acro{HARQ}{hybrid automatic repeat request}
 \acro{HBS}{home base station}
 \acro{HCN}{heterogeneous cellular network}
 \acro{HCS}{hierarchical cell structure}
  \acro{HD}{high definition}
 \acro{HDFP}{horizontal dynamic frequency planning}
 \acro{HE}{high efficiency}
 \acro{HeNB}{home eNodeB}
 \acro{HeNodeB}{home eNodeB}
 \acro{HetNet}{heterogeneous network}
 \acro{HiFi}{high fidelity}
 \acro{HII}{high interference indicator}
 \acro{HLR}{home location register}
 \acro{HMD}{head-mounted display}
 \acro{HNB}{home NodeB}
 \acro{HNBAP}{home NodeB application protocol}
 \acro{HNBGW}{home NodeB gateway}
 \acro{HNodeB}{home NodeB}
 \acro{HO}{handover}
 \acro{HOF}{handover failure}
 \acro{HOM}{handover hysteresis margin}
 \acro{HPBW}{half power beam width}
 \acro{HPLMN}{home PLMN}
 \acro{HPPP}{homogeneous Poison point process}
 \acro{HRD}{horizontal reflection diffraction}
 \acro{HSB}{hot spot boundary}
 \acro{HSDPA}{high speed downlink packet access}
 \acro{HSDSCH}{high-speed DSCH}
 \acro{HSPA}{high speed packet access}
 \acro{HSS}{home subscriber server}
 \acro{HSUPA}{high speed uplink packet access}
 \acro{HUA}{home user agent}
 \acro{HUE}{home user equipment}
 \acro{HVAC}{heating, ventilating, and air conditioning}
 \acro{IC}{interference cancellation}
 \acro{ICF}{initial control frame}
 \acro{ICI}{inter-carrier interference}
 \acro{ICIC}{intercell interference coordination}
 \acro{ICNIRP}{International Commission on Non-Ionising Radiation Protection}
  \acro{ICR}{initial control response}
 \acro{ICS}{IMS centralised service}
 \acro{ICT}{information and communication technology}
 \acro{ID}{identifier}
 \acro{IDFT}{inverse discrete Fourier transform}
 \acro{IE}{information element}
 \acro{IEEE}{Institute of Electrical and Electronics Engineers}
 \acro{IETF}{internet engineering task force}
 \acro{IFA}{inverted-F-antennas}
 \acro{IFFT}{inverse fast Fourier transform}
 \acro{i.i.d.}{independent and identical distributed}
 \acro{IIR}{infinite impulse response}
 \acro{IKE}{Internet key exchange}
 \acro{IKEv2}{Internet key exchange version 2}
 \acro{ILP}{integer linear programming}
 \acro{IMEI}{international mobile equipment identity}
 \acro{IMMW}{integrated mmWave}
 \acro{IMS}{IP multimedia subsystem}
 \acro{IMSI}{international mobile subscriber identity}
 \acro{IMT}{international mobile telecommunications}
 \acro{INH}{indoor hotspot}
 \acro{IOI}{interference overload indicator}
 \acro{IoT}{Internet of things}
 \acro{IP}{Internet protocol}
 \acro{IPSEC}{Internet protocol security}
 \acro{IR}{incremental redundancy}
 \acro{IRC}{interference rejection combining}
 \acro{ISD}{inter site distance}
 \acro{ISI}{inter symbol interference}
 \acro{ITU}{international telecommunication union}
 \acro{Iub}{UMTS interface between RNC and NodeB}
 \acro{IWF}{IMS interworking function}
 \acro{JFI}{Jain's fairness index}
 \acro{KPI}{key performance indicator}
 \acro{KNN}{$k$-nearest neighbours}
 \acro{L1}{layer one}
 \acro{L2}{layer two}
 \acro{L3}{layer three}
 \acro{LA}{location area}
 \acro{LAA}{licensed Assisted Access}
 \acro{LAC}{location area code}
 \acro{LAI}{location area identity}
 \acro{LAU}{location area update}
 \acro{LDA}{linear discriminant analysis} 
 \acro{LIDAR}{laser imaging detection and ranging}
 \acro{LLR}{log-likelihood ratio}
 \acro{LLS}{link-level simulation}
 \acro{LMDS}{local multipoint distribution service}
 \acro{LMMSE}{linear minimum mean-square-error}
 \acro{LoS}{line-of-sight}
 \acro{LPC}{logical PDCCH candidate}
 \acro{LPN}{low power node}
 \acro{LR}{likelihood ratio}
 \acro{LSAS}{large-scale antenna system}
 \acro{LSP}{large-scale parameter}
 \acro{LSTM}{long short term memory cell}
 \acro{LTE/SAE}{long term evolution/system architecture evolution}
 \acro{LTE}{long term evolution}
 \acro{LTE-A}{long term evolution advanced}
 \acro{LUT}{look up table}
 \acro{MAC}{medium access control}
 \acro{MaCe}{macro cell}
 \acro{MAE}{mean absolute error}
 \acro{MAP}{media access protocol}
 \acro{MAPC}{multi-access point coordination}
 \acro{MAPE}{mean absolute percentage error}
 \acro{MAXI}{maximum insertion}
 \acro{MAXR}{maximum removal}
 \acro{M-BA}{Multi-STA Block Ack}
 \acro{MBMS}{multicast broadcast multimedia service} 
 \acro{MBS}{macrocell base station}
 \acro{MBSFN}{multicast-broadcast single-frequency network}
 \acro{MC}{modulation and coding}
 \acro{MCB}{main circuit board}
 \acro{MCM}{multi-carrier modulation}
 \acro{MCP}{multi-cell processing}
 \acro{MCPA}{multi-carrier power amplifier}
 \acro{MCS}{modulation and coding scheme}
 \acro{MCSR}{multi-carrier soft reuse}
 \acro{MDAF}{management data analytics function}
 \acro{MDP}{markov decision process }
 \acro{MDT}{minimisation of drive tests}
 \acro{MEA}{multi-element antenna}
 \acro{MeNodeB}{Master eNodeB}
 \acro{MGW}{media gateway}
 \acro{MIB}{master information block}
 \acro{MIC}{mean instantaneous capacity}
 \acro{MIESM}{mutual information effective SINR mapping}
 \acro{MIMO}{multiple-input multiple-output}
 \acro{MINI}{minimum insertion}
 \acro{MINR}{minimum removal}
 \acro{MIP}{mixed integer program}
 \acro{MISO}{multiple-input single-output}
 \acro{ML}{machine learning}
 \acro{MLB}{mobility load balancing}
 \acro{MLD}{multi-link device} 
 \acro{STR}{simultaneous transmit and receive}
 \acro{MLMR}{multi-link multi-radio} 
 \acro{MLSR}{multi-link single-radio} 
 \acro{MM}{mobility management}
 \acro{MME}{mobility management entity}
 \acro{mMIMO}{massive multiple-input multiple-output}
 \acro{MMSE}{minimum mean square error}
 \acro{mMTC}{massive machine type communication}
 \acro{MNC}{mobile network code}
 \acro{MNO}{mobile network operator}
 \acro{MOS}{mean opinion score}
 \acro{MPC}{multi-path component}
 \acro{MR}{measurement report}
 \acro{MRC}{maximal ratio combining}
 \acro{MR-FDPF}{multi-resolution frequency-domain parflow}
 \acro{MRO}{mobility robustness optimisation}
 \acro{MRT}{maximum ratio transmission}
 \acro{MS}{mobile station}
 \acro{MSC}{mobile switching centre}
 \acro{MSE}{mean square error}
 \acro{MSISDN}{mobile subscriber integrated services digital network number}
 \acro{MTP}{motion-to-photon}
 \acro{MU}{multi-user}
 \acro{MUE}{macrocell user equipment}
 \acro{multi-RU}{multiple resource units}
 \acro{MU-MIMO}{multi-user multiple-input multiple-output}
 \acro{MU-RTS}{multi-user request-to-send}
 \acro{MVNO}{mobile virtual network operators}
 \acro{NACK}{negative acknowledgment}
 \acro{NAS}{non access stratum}
 \acro{NAV}{network allocation vector}
 \acro{NB}{Naive Bayes}   
 \acro{NCL}{neighbour cell list}
 \acro{NEE}{network energy efficiency}
  \acro{NF}{network function}
 \acro{NFV}{network functions virtualization}
 \acro{NG}{next generation}
 \acro{NGMN}{next generation mobile networks}
 \acro{NG-RAN}{next generation radio access network} 
 \acro{NIR}{non ionisation radiation}
 \acro{NLoS}{non-line-of-sight}
 \acro{NN}{nearest neighbour}
 \acro{NPCA}{non-primary channel access}
 \acro{NR}{new radio}
 \acro{NRMSE}{normalised root mean square error}
 \acro{NRTPS}{non-real-time polling service}
 \acro{NSS}{network switching subsystem}
 \acro{NTP}{network time protocol}
 \acro{NWG}{network working group}
 \acro{NWDAF}{network data analytics function} 
 \acro{OA}{open access}
 \acro{OAM}{operation, administration and maintenance}
 \acro{OC}{optimum combining}
 \acro{OCXO}{oven controlled oscillator}
 \acro{ODA}{omdi-directional antenna} 
 \acro{ODU}{optical distribution unit}
 \acro{OFDM}{orthogonal frequency division multiplexing}
 \acro{OFDMA}{orthogonal frequency division multiple access}
 \acro{OFS}{orthogonally-filled subframe}
 \acro{OLT}{optical line termination}
 \acro{OMP}{operating mode and parameter}
 \acro{ONT}{optical network terminal}
 \acro{OPEX}{operational expenditure}
 \acro{OSF}{odd subframe}
 \acro{OSI}{open systems interconnection}
 \acro{OSS}{operation support subsystem}
 \acro{OTT}{over the top}
 \acro{P2MP}{point to multi-point}
 \acro{P2P}{point to point}
 \acro{PAPR}{peak-to-average power ratio}
 \acro{PA}{power amplifier}
 \acro{PBCH}{physical broadcast channel}
 \acro{PC}{power control}
 \acro{PCB}{printed circuit board}
 \acro{PCC}{primary carrier component}
 \acro{PCCH}{paging control channel}
 \acro{PCCPCH}{primary common control physical channel}
 \acro{PCell}{primary cell}
 \acro{PCFICH}{physical control format indicator channel}
 \acro{PCH}{paging channel}
 \acro{PCI}{physical layer cell identity}
 \acro{PCPICH}{primary common pilot channel}
 \acro{PCPPH}{physical common packet channel}
 \acro{PDCCH}{physical downlink control channel}
 \acro{PDCP}{packet data convergence protocol}
 \acro{PDF}{probability density function}
 \acro{PDT}{proposed draft text}
 \acro{PDSCH}{physical downlink shared channel}
 \acro{PDU}{packet data unit}
 \acro{PeNB}{pico eNodeB}
 \acro{PeNodeB}{pico eNodeB}
 \acro{PER}{packet error rate}
 \acro{PF}{proportional fair}
 \acro{PGW}{packet data network gateway}
 \acro{PGFL}{probability generating functional}
 \acro{PhD}{doctor of philosophy}
 \acro{PHICH}{physical HARQ indicator channel}
 \acro{PHY}{physical}
 \acro{PIC}{parallel interference cancellation}
 \acro{PKI}{public key infrastructure}
 \acro{PL}{path loss}
 \acro{PMI}{precoding-matrix indicator}
 \acro{PLMN ID}{public land mobile network identity}
 \acro{PLMN}{public land mobile network}
 \acro{PML}{perfectly matched layer}
 \acro{PMF}{probability mass function}
 \acro{PMP}{point to multi-point}
 \acro{PN}{pseudorandom noise}
 \acro{POI}{point of interest}
 \acro{PON}{passive optical network}
 \acro{POP}{point of presence}
 \acro{PP}{point process}
 \acro{PPP}{Poisson point process}
 \acro{PPT}{PCI planning tools}
 \acro{PRACH}{physical random access channel}
 \acro{PRB}{physical resource block}
 \acro{PSC}{primary scrambling code}
 \acro{PSD}{power spectral density}
 \acro{PSS}{primary synchronisation channel}
 \acro{PSTN}{public switched telephone network}
 \acro{PTP}{point to point}
 \acro{PUCCH}{Physical Uplink Control Channel}
 \acro{PUE}{picocell user equipment}
 \acro{PUSC}{partial usage of subchannels}
 \acro{PUSCH}{physical uplink shared channel}
 \acro{QAM}{quadrature amplitude modulation}
 \acro{QCI}{QoS class identifier}
 \acro{QoE}{quality of experience}
 \acro{QoS}{quality of service}
 \acro{QPSK}{quadrature phase-shift keying}
 \acro{RA}{random access}
 \acro{RAB}{radio access bearer}
 \acro{RACH}{random access channel}
 \acro{RADIUS}{remote authentication dial-in user services}
 \acro{RAM}{Random Access Memory}
 \acro{RAN}{radio access network}
 \acro{RANAP}{radio access network application part}
 \acro{RAT}{radio access technology}
 \acro{RAU}{remote antenna unit}
 \acro{RAXN}{relay-aided x network}
 \acro{RB}{resource block}
 \acro{RCI}{re-establish cell id}
 \acro{RE}{resource efficiency}
 \acro{REB}{range expansion bias}
 \acro{REG}{resource element group}
 \acro{RF}{radio frequency}
  \acro{RFID}{radio frequency identification}
 \acro{RFP}{radio frequency planning}
 \acro{RI}{rank indicator}
 \acro{RL}{reinforcement learning}
 \acro{RLC}{radio link control}
 \acro{RLF}{radio link failure}
 \acro{RLM}{radio link monitoring}
 \acro{RMA}{rural macrocell}
 \acro{RMS}{root mean square}
 \acro{RMSE}{root mean square error}
 \acro{RN}{relay node}
 \acro{RNC}{radio network controller}
 \acro{RNL}{radio network layer}
 \acro{RNN}{recurrent neural network}
 \acro{RNP}{radio network planning}
 \acro{RNS}{radio network subsystem}
 \acro{RNTI}{radio network temporary identifier}
 \acro{RNTP}{relative narrowband transmit power}
 \acro{RPLMN}{registered PLMN}
 \acro{RPSF}{reduced-power subframes}
 \acro{RR}{round robin}
 \acro{RRC}{radio resource control}
 \acro{RRH}{remote radio head}
 \acro{RRM}{radio resource management}
 \acro{RS}{reference signal}
 \acro{RSC}{recursive systematic convolutional}
 \acro{RS-CS}{resource-specific cell-selection}
 \acro{RSQ}{reference signal quality}
 \acro{RSRP}{reference signal received power}
 \acro{RSRQ}{reference signal received quality}
 \acro{RSS}{reference signal strength}
 \acro{RSSI}{receive signal strength indicator}
 \acro{RTP}{real time transport}
 \acro{RTPS}{real-time polling service}
 \acro{RTS}{request-to-send}
 \acro{RTT}{round trip time}
  \acro{RU}{resource unit}
  \acro{RV}{random variable}
  \acro{RWP}{random waypoint}
 \acro{RX}{receive}
 \acro{S1-AP}{s1 application protocol}
 \acro{S1-MME}{s1 for the control plane}
 \acro{S1-U}{s1 for the user plane}
 \acro{SA}{simulated annealing}
 \acro{SACCH}{slow associated control channel}
 \acro{SAE}{system architecture evolution}
 \acro{SAEGW}{system architecture evolution gateway}
 \acro{SAIC}{single antenna interference cancellation}
 \acro{SAP}{service access point}
 \acro{SAR}{specific absorption rate}
 \acro{SARIMA}{seasonal autoregressive integrated moving average}
 \acro{SAS}{spectrum allocation server}
 \acro{SBS}{super base station}
 \acro{SCC}{standards coordinating committee}
 \acro{SCCPCH}{secondary common control physical channel}
 \acro{SCell}{secondary cell}
 \acro{SCFDMA}{single carrier FDMA}
 \acro{SCH}{synchronisation channel}
 \acro{SCM}{spatial channel model}
 \acro{SCN}{small cell network}
 \acro{SCOFDM}{single carrier orthogonal frequency division multiplexing}
 \acro{SCP}{single cell processing}
 \acro{SCTP}{stream control transmission protocol}
 \acro{SDCCH}{standalone dedicated control channel}
 \acro{SDMA}{space-division multiple-access}
  \acro{SDO}{standard development organization}
 \acro{SDR}{software defined radio}
 \acro{SDU}{service data unit}
 \acro{SE}{spectral efficiency}
 \acro{SeNodeB}{secondary eNodeB}
 \acro{SFBC}{space frequency block coding}
 \acro{SFD}{specification framework document}
 \acro{SFID}{service flow ID}
 \acro{SG}{study group}
 \acro{SGSN}{serving GPRS support node}
 \acro{SGW}{serving gateway}
 \acro{SHAP}{shapley additive explanations}
 \acro{SI}{system information}
 \acro{SIB}{system information block}
 \acro{SIB1}{systeminformationblocktype1}
 \acro{SIB4}{systeminformationblocktype4}
 \acro{SIC}{successive interference cancellation}
 \acro{SIFS}{short interframe space}
 \acro{SIGTRAN}{signalling transport}
 \acro{SIM}{subscriber identity module}
 \acro{SIMO}{single input multiple output}
 \acro{SINR}{signal to interference plus noise ratio}
 \acro{SIP}{session initiated protocol}
 \acro{SIR}{signal to interference ratio}
 \acro{SISO}{single input single output}
 \acro{SLAC}{stochastic local area channel}
 \acro{SLL}{secondary lobe level}
 \acro{SLNR}{signal to leakage interference and noise ratio}
 \acro{SLO}{single-link operation}
 \acro{SLS}{sector level sweep}
  \acro{SMB}{small and medium-sized businesses}
 \acro{SmCe}{small cell}
 \acro{SMS}{short message service}
 \acro{SN}{serial number}
 \acro{SNMP}{simple network management protocol}
 \acro{SNR}{signal to noise ratio}
 \acro{SOCP}{second-order cone programming}
 \acro{SOHO}{small office/home office}
 \acro{SON}{self-organising network}
 \acro{son}{self-organising networks}
 \acro{SOT}{saving of transmissions}
 \acro{SPS}{spectrum policy server}
 \acro{SRCON}{simulated reality of communication networks} 
 \acro{SRS}{sounding reference signals}
 \acro{SS}{synchronization signal}
 \acro{SSB}{synchronization Signal/PBCH block}
 \acro{SIB1}{system information block 1} 
 \acro{SSL}{secure socket layer}
 \acro{SSMA}{spread spectrum multiple access}
 \acro{SSS}{secondary synchronisation channel}
 \acro{STA}{station}
 \acro{STBC}{space-time block coding}
 \acro{SUI}{stanford university interim}
 \acro{SVR}{support vector regression}
 \acro{TA}{timing advance}
 \acro{TAC}{tracking area code}
 \acro{TAI}{tracking area identity}
 \acro{TAS}{transmit antenna selection}
 \acro{TAU}{tracking area update}
 \acro{TB}{trigger-based}
 \acro{TCH}{traffic channel}
 \acro{TCO}{total cost of ownership}
 \acro{TCP}{transmission control protocol}
 \acro{TCXO}{temperature controlled oscillator}
 \acro{TD}{timeout discards}
 \acro{TDD}{time division duplexing}
 \acro{TDM}{time division multiplexing}
 \acro{TDMA}{time division multiple access}
 \acro{TDoA}{time difference of arrival}
 \acro{TEID}{tunnel endpoint identifier}
 \acro{TF}{trigger frame}
 \acro{TG}{task group}
 \acro{TGbq}{802.11bq task group}
 \acro{TLS}{transport layer security}
 \acro{TNL}{transport network layer}
 \acro{ToA}{time of arrival}
 \acro{TP}{throughput}
 \acro{TPC}{transmit power control}
 \acro{TPM}{trusted platform module}
 \acro{TR}{transition region}
 \acro{TRN}{training}
 \acro{TRX}{transceiver}
 \acro{TS}{tabu search}
 \acro{TSG}{technical specification group}
 \acro{TTG}{transmit/receive transition gap}
 \acro{TTI}{transmission time interval}
 \acro{TTT}{time-to-trigger}
 \acro{TU}{typical urban}
 \acro{TV}{television}
 \acro{TWXN}{two-way exchange network}
 \acro{TX}{transmit}
 \acro{TXS}{TXOP sharing}
 \acro{UARFCN}{UTRA absolute radio frequency channel number}
 \acro{UAV}{unmanned aerial vehicle}
 \acro{UCI}{uplink control information}
 \acro{UDP}{user datagram protocol}
 \acro{UDN}{ultra-dense network}
 \acro{UE}{user equipment}
 \acro{UGS}{unsolicited grant service}
 \acro{UHD}{ultra-high definition}
 \acro{UICC}{universal integrated circuit card}
 \acro{UK}{united kingdom}
 \acro{UL}{uplink}
 \acro{UMA}{unlicensed mobile access}
 \acro{UMi}{urban micro}
 \acro{UMTS}{universal mobile telecommunication system}
 \acro{UN}{United Nations}
 \acro{URLLC}{ultra-reliable low-latency communication}
 \acro{US}{upstream}
 \acro{USIM}{universal subscriber identity module}
 \acro{UTD}{theory of diffraction}
 \acro{UTRA}{UMTS terrestrial radio access}
 \acro{UTRAN}{UMTS terrestrial radio access network}
 \acro{UWB}{ultra wide band}
 \acro{VBR}{variable bit rate}
 \acro{VD}{Vsync discards}
 \acro{VDFP}{vertical dynamic frequency planning}
 \acro{VDSL}{very-high-bit-rate digital subscriber line}
 \acro{VeNB}{virtual eNB}
 \acro{VeNodeB}{virtual eNodeB}
 \acro{VF-RTT}{video frame round-trip time}
 \acro{VIC}{victim cell}
 \acro{VLR}{visitor location register}
 \acro{VNF}{virtual network function}
 \acro{VoIP}{voice over IP}
 \acro{VoLTE}{voice over LTE}
 \acro{VPLMN}{visited PLMN}
 \acro{VR}{virtual reality}
 \acro{VRAN}{virtualized radio access network}
 \acro{VSync}{vertical synchronization} 
 \acro{WCDMA}{wideband code division multiple access}
 \acro{WEP}{wired equivalent privacy}
 \acro{WG}{working group}
 \acro{WHO}{world health organisation}
 \acro{Wi-Fi}{Wi-Fi}
 \acro{WiMAX}{wireless interoperability for microwave access}
 \acro{WiSE}{wireless system engineering}
 \acro{WLAN}{wireless local area network}
 \acro{WMAN}{wireless metropolitan area network}
 \acro{WNC}{wireless network coding}
 \acro{WRAN}{wireless regional area network}
 \acro{WSEE}{weighted sum of the energy efficiencies}
 \acro{WPEE}{weighted product of the energy efficiencies}
 \acro{WMEE}{weighted minimum of the energy efficiencies}
 \acro{X2}{x2}
 \acro{X2-AP}{x2 application protocol}
 \acro{ZF}{zero forcing}
 \acro{UHR}{ultra high reliability}
 \acro{PAR}{project authorization request}
 \acro{MPDU}{MAC protocol data unit}
 \acro{A-MPDU}{aggregate MAC protocol data unit}
 \acro{BSS}{basic service set}
 \acro{MLO}{multi-link operation} 
 \acro{OBSS}{overlapping basic service set} 
\acro{TXOP}{transmission opportunity}
\acrodefplural{TXOP}{transmission opportunities}
 \acro{Co-BF}{coordinated beamforming}
 \acro{R-TWT}{restricted target wake time}
 \acro{TWT}{target wake time}
 \acro{SP}{service period}
 \acro{GI}{guard interval}
 \acro{NDP}{null data packet}
 \acro{SR}{spatial reuse}
 \acro{Co-SR}{coordinated spatial reuse}
 \acro{DRU}{distributed resource unit}
 \acro{RRU}{regular resource unit}
 \acro{DBW}{distribution bandwidth}
 \acro{LTF}{long training field}
 \acro{CSD}{cyclic shift diversity}
 \acro{ELR}{enhanced long range}
 \acro{DSSS}{direct sequence spread spectrum} 
 \acro{UEQM}{unequal modulation}
 \acro{LDPC}{low-density parity check} 
 \acro{IM}{interference mitigation} 
 \acro{XR}{extended reality}
 \acro{QC-LDPC}{quasi-cyclic LDPC}
 \acro{PPDU}{physical layer protocol data unit}
\end{acronym}

\end{document}